\documentclass[aps,twocolumn,superscriptaddress,nofootinbib,a4paper,longbibliography]{revtex4-2}

\usepackage{times}
\usepackage{epsfig}
\usepackage{amsfonts}
\usepackage{amsmath}
\usepackage{amsthm}
\usepackage{amssymb}					
\usepackage{amsthm}
\usepackage{dsfont}
\usepackage{bm}
\usepackage{mathtools}
\usepackage{color}
\usepackage{multirow}
\usepackage[normalem]{ulem}
\newcommand{\stkout}[1]{\ifmmode\text{\sout{\ensuremath{#1}}}\else\sout{#1}\fi}
\usepackage{latexsym}
\usepackage{mathrsfs}
\usepackage{natbib}
\usepackage{verbatim}
\usepackage[T1]{fontenc}
\usepackage{float}
\usepackage{graphicx}
\usepackage{xcolor}
\usepackage{physics}
\usepackage{soul}

\newtheorem{theorem}{Theorem}

\newtheorem{definition}[theorem]{Definition}

\usepackage[colorlinks=true,linkcolor=blue,citecolor=magenta,urlcolor=blue]{hyperref}

\begin{document}

%%%%%%%%%%%%%%%%%%%%%%%%%%%%%%%%%%%%%%%%%%%%%%%%%%%%%%%%%%%%%%%%%%%

\title{Non-projective Bell state measurements}

\author{Amanda Wei}
\affiliation{Department of Physics and NanoLund, Lund University, Box 118, 22100 Lund, Sweden}

\author{Gabriele Cobucci}
\affiliation{Department of Physics and NanoLund, Lund University, Box 118, 22100 Lund, Sweden}

\author{Armin Tavakoli}
\affiliation{Department of Physics and NanoLund, Lund University, Box 118, 22100 Lund, Sweden}

\begin{abstract}
	The  Bell state measurement (BSM) is the projection of two qubits onto four orthogonal maximally entangled states. Here, we first propose how to appropriately define more general BSMs, that have more than four possible outcomes, and then study whether they exist in quantum theory. We observe that non-projective BSMs can be defined in a systematic way in terms of equiangular tight frames of maximally entangled states, i.e.~a set of maximally entangled states, where every pair is equally, and in a sense maximally, distinguishable. We show that there exists a five-outcome BSM through an explicit construction, and find that it admits a simple geometric representation. Then, we prove that there exists no larger BSM on two qubits by showing that no six-outcome BSM is possible. We also determine the most distinguishable set of six equiangular maximally entangled states and show that it falls only somewhat short of  forming a valid quantum measurement. Finally, we study the non-projective BSM in the contexts of both local state discrimination and entanglement-assisted quantum communication.  Our results put forward natural forms of non-projective joint measurements and provide insight on the geometry of entangled quantum states. 
\end{abstract}

\date{\today}

\maketitle

\section{Introduction}
The Bell state measurement  (BSM) is a  projection of a pair of qubits onto an orthonormal basis of maximally entangled two-qubit states. It plays a key role in quantum information science and it is at the heart of several paradigmatic protocols, such as  teleportation, dense coding and entanglement swapping.  However, many interesting quantum measurements are not projective. For instance, while  projective measurements have at most as many outcomes as the Hilbert space dimension, $d$, non-projective measurements can have up to $d^2$ outcomes \cite{Ariano2005}. Such measurements have many known applications, in e.g.~state discrimination \cite{Ivonovic1981}, tomography \cite{Caves2002}, entanglement detection \cite{Shang2018, Morelli2023, Morelli2024} and device-independence \cite{Acin2016, Tavakoli2021}. Realising them has been the focus of many experiments, on qubits \cite{Bian2015, Gomez2016, Tavakoli2020c} and beyond \cite{Martínez2023, Feng2023, Wang2023}. 

Here, we ask an elementary conceptual question about joint measurements in quantum theory: does there exist a non-projective counterpart of the standard BSM? This is not only interesting for understanding quantum theory and nature of relevant non-projective joint measurements, but also to understand the types of geometries admitted by quantum states, see e.g.~the book \cite{BengtssonZyczkowski2006}. The latter has catalysed much research on both exotic quantum states and measurements \cite{Renes2004, Zauner2011, Wiesniak2011, Gisin2019, Stacey2020, Tavakoli2020, Czartowski2020, Tavakoli2020b, Geng2021}.

We propose a natural definition of non-projective BSMs by identifying them with a mathematical object called an equiangular tight frame (ETF), better known to physicists through the famous special instance called a symmetric informationally complete POVM \cite{Renes2004, Zauner2011}. An ETF is an overcomplete basis with the additional property that all constitutent pairs of vectors are separated by the same angle. To interpret this as a Bell state measurement, we will also require that all vectors are maximally entangled. In this picture, the standard BSM is the trivial special case. With the definition in hand, the natural question is whether non-projective BSMs are at all possible in quantum theory. We answer this in the positive by explicitly constructing a five-outcome BSM. In contrast to the standard BSM, this non-projective BSM appears to admit no group structure. Nevertheless, it can be represented in an elegant way on the Bloch sphere. Next, we set out to determine the ``largest'' possible non-projective BSM. We  prove that no more than five outcomes are possible. This is achieved through an impossibility proof for a six-outcome BSM. To understand this no-go result in terms of Hilbert space geometry, we also determine the largest possible separation angle between six equiangular maximally entangled states and find that it is only somewhat smaller than the critical angle necessary to form a non-projective BSM. Subsequently, we comparatively study the standard BSM and our non-projective BSM via the task of local state discrimination. We find that the latter measurement is more nonlocal, in the sense that the relative performance of local and global measurements to distinguish the associated Bell states is lower, especially when two copies of the state are available. Finally, we consider the entanglement-assisted prepare-and-measure scenario \cite{Tavakoli2021b, Pauwels2022} and show that there exists a simple task in which state geometry associated with the non-projective BSM leads to the optimal quantum protocol.

\section{Preliminaries}
Consider a set of $N\geq m$  pure states, $\mathcal{S}=\{\ket{\psi_i}\}_{i=1}^N$, on $\mathbb{E}^m$, where we choose either the real or the complex space $\mathbb{E}\in\{\mathbb{R},\mathbb{C}\}$. The set $\mathcal{S}$ is called a tight frame if for every state $\ket{\phi}\in\mathbb{E}^m$ it holds that $\sum_{i=1}^N \left|\braket{\phi}{\psi_i}\right|^2=a$, for some constant $a$. This can be equivalently stated as $\tr(S\ketbra{\phi})=a$ $\forall \phi$, where  $S=\sum_{i=1}^N \ketbra{\psi_i}$ is called the frame operator. A tight frame is hence equivalent to having $S=a\openone$. This permits us to interpret a tight frame as a quantum measurement; the set $\{\frac{1}{a}\ketbra{\psi_i}\}_{i=1}^N$ forms an $N$-outcome  POVM. 

The set $\mathcal{S}$ is called equiangular if for all $i\neq j$,
\begin{equation}\label{equiangular}
\left|\braket{\psi_i}{\psi_j}\right|^2 = \alpha.
\end{equation}
Thus, when $\mathcal{S}$ is both equiangular and a tight frame, we call it an ETF. In an ETF, the overlap constant $\alpha$ must attain the value  $\alpha^*=\frac{N-m}{m(N-1)}$. This follows from the Welch bound \cite{Welch1974}, which states that for any set of states, 
\begin{equation}\label{Welch}
\max_{i\neq j} \left|\braket{\psi_i}{\psi_j}\right|^2 \geq \alpha^*,
\end{equation}
with equality if and only if $\mathcal{S}$ is an ETF. For completeness, we reproduce a proof of this statement in Appendix~\ref{AppWelch}. Therefore, for a given $(N,m)$, an ETF (if it exists) can be interpreted as a set of states in which all pairs have the same angle and where that angle saturates the bound stipulated by the Welch bound.  Note also that the ETF condition determines the frame constant, $a$, because  $\tr(S^2)=N+N(N-1)\alpha^* =a^2 m$, which implies  $a=\frac{N}{m}$.

What is the largest possible number of states, $N_\text{max}$, in an ETF of dimension $m$? For both real-valued and complex-valued quantum states, the answer is unknown but the Gerzon bound limits both cases \cite{Lemmens1973}; $N_\text{max}\leq \frac{m(m+1)}{2}$ for real spaces and $N_\text{max}\leq m^2$ for complex spaces. The former inequality is rarely saturated \cite{Stacey2020} but the latter inequality is believed to always be possible to saturate \cite{Zauner2011}. Proving the latter claim is equivalent to the open problem of finding a SIC-POVM in every dimension \cite{Horodecki2022}.

\section{Non-projective Bell state measurements}
We are now ready to define non-projective BSMs in terms of maximally entangled ETFs. 
\begin{definition}
	An $N$-outcome Bell state measurement on a Hilbert space $\mathbb{C}^d\otimes \mathbb{C}^d$ is a POVM $\{\frac{d^2}{N}\ketbra{\psi_i}\}_{i=1}^N$ where $N\geq d^2$, all states $\ket{\psi_i}$ are maximally entangled and the set $\{\ket{\psi_i}\}$ is an ETF.   
\end{definition}
Notice that the tight frame property guarantees that the interpretation as a quantum measurement is valid. The equiangular property guarantees that all outcome-operators are symmetric w.r.t each other, i.e.~the labeling of the maximally entangled states is irrelevant. The maximal entanglement guarantees that all states $\ket{\psi_i}$ are unitarily equivalent to the standard maximally entangled state $\ket{\phi^+}=\frac{1}{\sqrt{d}}\sum_{i=0}^{d-1}\ket{ii}$. The latter follows from the Choi isomorphism between maximally entangled states and unitary channels, i.e.~every maximally entangled state can be written as
\begin{equation}\label{maxent}
\ket{\psi_i}=U_i\otimes \openone\ket{\phi^+},
\end{equation}
for some $d$-dimensional unitary $U_i$. The saturation of the Welch bound \eqref{Welch} can then be expressed only in terms of local unitary operators, i.e.~$\forall i\neq j$
\begin{equation}\label{tounitary}
\left|\braket{\psi_i}{\psi_j}\right|^2=\frac{1}{d^2}\left|\tr(U_i^\dagger U_j)\right|^2=\alpha^*.
\end{equation}
It is instructive to note that the standard BSM, corresponding to $N=d^2$ orthogonal maximally entangled states, is a trivial special case of this framework, because it gives $\alpha^*=0$ which corresponds to perfectly distinguishable states. 

In what follows we will focus on non-projective BSMs for two qubits ($d=2$). The analysis can be much simplified by  making use of an isomorphism between the set of two-qubit maximally entangled states and pure quantum states in $\mathbb{R}^4$.

\section{Mapping the problem to real spaces}
It is well-known that the group of rotations in four-dimensional Euclidean space, $\text{SO}(4)$, is isomorphic to two copies of the qubit special unitary group, $\text{SU}(2)$. In a similar way, the set of maximally entangled two-qubit states is in one-to-one correspondence with the set of real-valued four-dimensional states \cite{Pauwels2022}. 

Concretely,  $\ket{\psi}=(x_1,x_2,x_3,x_4)^T\equiv \bar{x}\in \mathbb{R}^4$ can be mapped to the qubit unitary, $\bar{x}\rightarrow ix_1\openone+x_2\sigma_X+x_3\sigma_Y+x_4\sigma_Z=U$ where $\{\sigma_X,\sigma_Y,\sigma_Z\}$ are the Pauli matrices. The opposite direction is achieved by selecting any $U\in \text{SU}(2)$ and associating the state vector $U\rightarrow \frac{-i}{2}\left(\tr(iU),\tr(U\sigma_X),\tr(U\sigma_Y),\tr(U\sigma_Z)\right)=\bar{x}$.  If two states  $\ket{\psi},\ket{\varphi}\in\mathbb{R}^4$ have coordinates $\bar{x}$ and $\bar{y}$ respectively, we have $\braket{\psi}{\varphi}=\bar{x}\cdot \bar{y}$. Letting $U_\psi$ and $U_\varphi$ be the associated unitaries, we obtain $\Tr(U_\psi^\dagger U_\varphi)=2(\bar{x}\cdot \bar{y})$. Inserted into \eqref{tounitary}, we see that the Welch bound is saturated for the qubit unitaries if and only if it is saturated for the real states in $\mathbb{R}^4$.

This is a useful connection because, in contrast to the space of maximally entangled states, ETFs in Euclidean space have been studied in the mathematics literature. Importantly, it has been shown that there exists no set of seven equiangular states on $\mathbb{R}^4$ (with $\alpha<1$) \cite{Haantjes1948}. This  implies that no set of maximally entangled states with $N\geq 7$ elements is possible, and hence also that no non-projective BSM with $N\geq 7$ outcomes exists. In Appendix~\ref{AppIcosa}, we apply the isomorphism to the six equiangular states in $\mathbb{R}^4$ proposed in \cite{Lemmens1973}, based on the regular icosahedron in $\mathbb{R}^3$, to identify a corresponding equiangular set of maximally entangled two-qubit states. However,  this construction is far from saturating the Welch bound \eqref{Welch} and it is therefore no non-projective BSM.

\section{Five-outcome Bell state measurement}
We now construct a five-outcome BSM for two qubits. To derive it, we have considered five generic states in $\mathbb{R}^4$, imposed the equiangular condition \eqref{equiangular} with $\alpha=\alpha^*$ and then solved the resulting system of equations. Then, we use the above isomorphism to transform the solution to qubit unitaries  $\{U_i\}$, which corresponds to a set of maximally entangled states via Eq.~\eqref{maxent}. The solution takes the form
\begin{align}\label{bsm5states}\nonumber
|\psi_{1}\rangle &= \frac{i}{\sqrt{2}}(\Phi^{+}-\Psi^{-}) \\\nonumber
|\psi_{2}\rangle &= \frac{1}{\sqrt{2}}(ai\Phi^{+}+d\Phi^{-}+b\Psi^{+}-ci\Psi^{-}), \\\nonumber
|\psi_{3}\rangle &= \frac{1}{\sqrt{2}}(ai\Phi^{+}-d\Phi^{-}-b\Psi^{+}-ic\Psi^{-}),\\\nonumber
|\psi_{4}\rangle &= \frac{1}{\sqrt{2}}(ci\Phi^{+}+b\Phi^{-}-d\Psi^{+}-ai\Psi^{-}),\\
|\psi_{5}\rangle &= \frac{1}{\sqrt{2}}(ci\Phi^{+}-b\Phi^{-}+d\Psi^{+}-ai\Psi^{-}),
\end{align}
where $ \Phi^\pm=\frac{1}{\sqrt{2}}\left(\ket{00}\pm \ket{11}\right)$ and  $\Psi^\pm=\frac{1}{\sqrt{2}}\left(\ket{01}\pm \ket{10}\right)$ are the standard Bell states. The constants are given by $a = \frac{-1+\sqrt{5}}{4}$, $b = \sqrt{\frac{5+\sqrt{5}}{8}}$, $c = -\frac{1+\sqrt{5}}{4}$ and $d = \sqrt{\frac{5-\sqrt{5}}{8}}$. One can straightforwardly verify that $\left|\braket{\psi_i}{\psi_j}\right|^2=\frac{1}{16}$, which saturates the Welch bound  for $N=5$ and $m=d^2=4$. Hence, $\{\frac{4}{5}\ketbra{\psi_i}\}_{i=1}^5$ is a five-outcome non-projective BSM. The unitaries \eqref{maxent} corresponding to this construction are listed in Appendix~\ref{AppUnitaries}.

The non-projective BSM is unique up to local unitaries. To see this, let $V$ be the $d\times N$ matrix whose columns are $\bar{x}_i$. Let $G_{i,j}=\bar{x}_i\cdot \bar{x}_j$ be the $N\times N$ Gram matrix for the states in $\bar{x}_i\in\mathbb{R}^d$. Let also $S=\sum_{i=1}^N \bar{x}_i \otimes \bar{x}_j^T$ be the frame operator. Equivalently, they are expressed in terms of $V$ as $G=V^\dagger V$ and $S=VV^\dagger$. Then, $G$ and $S$ share the same eigenvalues. To see this, if $\lambda$ is an eigenvalue of $G$ then $G \bar{z}=V^\dagger V\bar{z}=\lambda \bar{z}$. Thus, $V(V^\dagger V)\bar{z}=V(\lambda \bar{z})$ which is equivalent to $S(V\bar{z})=\lambda(V\bar{z})$. Now if two Gram matrices $G_{1}$ and $G_{2}$ represent two ETFs of $N$ vectors in $\mathbb{R}^{d}$, then they must have the same characteristic polynomial, $\lambda^{N-d}(\lambda-N/d)^{d}$. This follows from the fact that the associated frame operators are $S_1=S_2=\frac{N}{d}\openone$.  From \cite{vanLint1966}, any two Gram matrices with the same characteristic polynomials can be obtained from one another through a sequence of unitary operations. Thus, we conclude that $G_{1}=UG_{2}U^{\dagger}$, from which we can relate the vectors of the equiangular lines explicitly, as given by $V_{1}=V_{2}U^\dagger$.

To better understand the non-projective BSM, we can picture the unitaries on the Bloch sphere. Every qubit unitary can, up to a global phase, be written as
\begin{equation}\label{unitary}
U=\cos\gamma \hspace{1mm}\openone +i\sin\gamma\hspace{1mm} \vec{n}\cdot\vec{\sigma},
\end{equation}
where $\gamma$ is the rotation angle, $\vec{n}$ is a normalised Bloch vector representing the axis of rotation and $\vec{\sigma}=(\sigma_X,\sigma_Y,\sigma_Z)$. To simplify the description, we first  rotate all the states \eqref{bsm5states} with the local unitary, $U_1^\dagger$, that maps $\ket{\psi_1}$ to $\ket{\phi^+}$. The set of states now corresponds to the unitaries $\{\openone,\{V_i\}_{i=2}^5\}$ where $V_i=U_1^\dagger U_i$. Representing the four unitaries $\{V_i\}$ as in Eq.~\eqref{unitary}, we find that they all correspond to the same rotation angle, $\gamma=\arccos(1/4)$, and that their respective rotation axes form the vertices of a regular tetrahedron inside the Bloch sphere. 

Furthermore, we have studied whether five-outcome two-qubit measurements are possible also when all the states are equally but not maximally entangled. This amounts to states of the form $V_i\otimes W_i\ket{\psi_\theta}$, where $V_i$ and $W_i$ are local unitaries and $\ket{\psi_\theta}=\cos\theta\ket{00}+\sin\theta\ket{11}$ for some $\theta\in[0,\frac{\pi}{4})$. The four-outcome (basis) case was studied in \cite{Pimpel2023}. We selected several values of $\theta$ and could always numerically find an ETF. The search can be performed efficiently by first parameterising all the single-qubit unitaries on the Bloch sphere, then definining the function $f=\sum_{j=1}^N\sum_{k=j+1}^N \big(\frac{1}{4}\tr\big(U^\dagger_jU_k\big)^2-z\big)^2$ and finally minimising $z$ under the constraint $f\leq \epsilon$, where we choose $\epsilon$ positive and very close to zero. If this, within tolerance, returns $z=\alpha^*$, then it corresponds to the desired measurement. In particular, for $\theta=0$, corresponding to a product state, we find that the five-outcome POVM admits a simple expression; we give the corresponding unitaries $\{V_i\otimes W_i\}_{i=1}^5$ explicitly in Appendix~\ref{AppProduct}.

\section{No six-outcome Bell state measurement}
We now show that there exists no six-outcome BSM. Our argument is based on directly leveraging the $\text{SU}(2)$ parameterisation of maxmially entangled two-qubit states and then use the Bloch representation \eqref{unitary} of the unitaries, namely $U_i=\cos\gamma_i\hspace{1mm}\openone+i\sin\gamma_i \hspace{1mm}\vec{n}_i\cdot 
\vec{\sigma}$, with $i=0,1,\ldots,5$. It then follows that
\begin{equation}\label{step}
\tr\left(U_i^\dagger U_j\right)=2\left(\cos\gamma_i\cos\gamma_j+\sin\gamma_i\sin\gamma_j\hspace{1mm}\vec{n}_i\cdot\vec{n}_j\right),
\end{equation}
which is notably a real number. W.l.g~we can take the first unitary to be $U_0=\openone$, i.e.~$\gamma_0=0$. The equiangular condition in Eq.~\eqref{tounitary} then implies that $\cos\gamma_i=\pm \sqrt{\alpha^*}$ for $i=1,\ldots,5$, i.e.~the rotation angle is fixed (up to a sign) for all the unitaries. Introduce the notation $\cos\gamma_i=(-1)^{c_i}|\cos\gamma_i|$ and $\sin\gamma_i=(-1)^{s_i}|\sin\gamma_i|$, where $c_i,s_i\in\{0,1\}$. Inserting this into \eqref{step}, saturating the Welch bound means $\forall i\neq j$,
\begin{multline}
\alpha^*(\alpha^*-1) +(\vec{n}_i\cdot\vec{n}_j)^2 (1-\alpha^*)^2\\
-2(-1)^{t_i+t_j}\alpha^*(1-\alpha^*)\vec{n}_i\cdot\vec{n}_j=0,
\end{multline}
where $t_i=c_i+s_i$ and $t_j=c_j+s_j$. Here, $\alpha^*=\frac{1}{10}$ is the Welch bound \eqref{Welch} for $N=6$ and $m=d^2=4$. Solving this equation in $\vec{n}_i\cdot\vec{n}_j$ gives
\begin{equation}\label{conds}
\vec{n}_i\cdot\vec{n}_j=\frac{(-1)^{t_i+t_j}\alpha^*\pm \sqrt{\alpha^*}}{1-\alpha^*}.
\end{equation}
To see if these conditions can be satisfied, we use spherical symmetry to w.l.g choose the Bloch vectors of the second and third unitaries as $\vec{n}_1=(0,0,1)$ and  $\vec{n}_2=(\sin\theta_2,0,\cos\theta_2)$. The remaining  three unitaries must be given arbitrary Bloch vectors $\vec{n}_i=\left(\sin\theta_i\cos\phi_i,\sin\theta_i \sin\phi_i,\cos\theta_i\right)$, for $i=3,4,5$. Hence, $\vec{n}_1\cdot\vec{n}_i=\cos\theta_i$ 
for $i=2,\ldots,5$, and $\vec{n}_2\cdot\vec{n}_i=\sin\theta_2\sin\theta_i\cos\phi_i+\cos\theta_2\cos\theta_i$ for $i=3,4,5$. The first equation determines (up to a sign) $\theta_i$ and then inserted into the second equation, it determines (up to a sign) $\phi_i$. We have considered every sign combination, for every choice of $\{t_i,t_j\}_{i,j}$, and verified that the equations \eqref{conds} are never satisfied. Hence, no six-outcome BSM exists.

\section{Optimal  separation for six equiangular maximally entangled states}
While no six-outcome BSM is possible, we have previously seen that six equiangular maximally entangled states can be constructed via the vertices of the regular icosahedron. This leads to a natural question: what is the largest separation angle possible between six equiangular maximally entangled states? In other words, how ``close'' can we come to the critical separation angle necessary for a non-projective BSM? While the Welch bound stands at $\alpha^*=\frac{1}{10}$, we show that the smallest possible overlap value is $\alpha=\frac{1}{9}$. 

First, we give an explicit construction of the states $\{|\psi_i\rangle\}_{i=1}^6$ which achieve $\alpha=\frac{1}{9}$. The states are
\begin{align}
\begin{split}
|\psi_{1}\rangle&=\Phi^{+}\\
|\psi_{2}\rangle&=-\frac{1}{3}\Phi^{+}+i\frac{2\sqrt{2}}{3}\Phi^{-}\\
|\psi_{3}\rangle&=\frac{1}{3}\Phi^{+}-i\frac{\sqrt{2}}{6}\Phi^{-}-i\sqrt{\frac{5}{6}}\Psi^{+} \\
|\psi_{4}\rangle &=\frac{1}{3}\Phi^{+}+i\frac{\sqrt{2}}{3}\Phi^{-}+i\sqrt{\frac{2}{15}}\Psi^{+}+2\sqrt{\frac{2}{15}}\Psi^{-} \\
|\psi_{5}\rangle &= \frac{1}{3}\Phi^{+}+i\frac{\sqrt{2}}{3}\Phi^{-}-i\sqrt{\frac{2}{15}}\Psi^{+}-2\sqrt{\frac{2}{15}}\Psi^{-} \\
|\psi_{6}\rangle &= \frac{1}{3}\Phi^{+}-i\frac{\sqrt{2}}{6}\Phi^{-}+i\frac{3}{\sqrt{30}}\Psi^{+}-\frac{4}{\sqrt{30}}\Psi^{-}
\end{split}
\end{align}
The associated qubit unitaries are given in Appendix~\ref{AppUnitaries2}.

Second, we must show that no smaller value of $\alpha$ is possible. To this end, we use the isomorphism between maximally entangled states and pure states in $\mathbb{R}^4$. Recall that the Gram matrix $G$ and frame operator $S$ have the same eigenvalues. Since $S$ is a $d\times d$ matrix, we have that $\text{rank}(G)\leq d$. Thus, the characteristic polynomial of $G$ must have  at least $N-d$ zero-eigenvalues. Recall also that $G\succeq 0$ by definition. Furthermore, for equiangular states, all off-diagonals of $G$ are $\pm\sqrt{\alpha}$. Since $G$ is symmetric and we have $d=4$, it has $15$ non-trivial off-diagonals, corresponding to $2^{15}$ possible sign combinations. A simpler approach than checking them all as a function of $\alpha$, is to notice that the determinant $\det(G-\lambda \openone)$ is invariant under multiplications of the $k$'th row and column by $-1$. Up to this degree of freedom, it has been shown that the set of Gram matrices for equiangular states can be classified into just 16 classes \cite{vanLint1966}. In Appendix~\ref{AppPolynomials}, we compute  the 16 characteristic polynomials. By inspecting them, it can be shown that the smallest $\alpha$ compatible with the positivity- and rank-constraints of $G$ is $\alpha=\frac{1}{9}$.

\section{Local state discrimination} We now comparatively study the nonlocality of the four- and five-outcome BSMs via the quantum information task of local state discrimination. A source randomly emits a state selected from a set $\mathcal{S}=\{\rho_i\}$ and the task is to guess the classical label by making a measurement. The optimal average success probability is 
\begin{equation}
P_t(\mathcal{S})=\max_{M\in t}\frac{1}{|\mathcal{S}|}\sum_{i=1}^{|\mathcal{S}|} \tr(\rho_i M_i),
\end{equation}
where $\{M_i\}$ is a measurement restricted to a certain class, labeled $t$. Commonly considered classes are $t\in\{\text{LO}, \text{LOCC}, \text{PPT},\text{G}\}$, corresponding local operations (LO), LO with classical communication (LOCC), measurements with a positive partial transpose and global measurements.  For the Bell states,  $\mathcal{S}=\{\Phi^+,\Phi^-,\Psi^+,\Psi^-\}$, it holds that  $P_\text{LO}=P_\text{LOCC}=P_\text{PPT}=\frac{1}{2}$ \cite{Ghosh2001, Bandyopadhyay2013}, which is saturated by  measuring $\sigma_Z\otimes \sigma_Z$ and guessing on $\Phi^+$ ($\Psi^+$) when the outcome is zero (one).  In contrast, if two copies of the Bell states are provided, i.e.~$\mathcal{S}=\{(\Phi^+)^{\otimes2},(\Phi^-)^{\otimes2},(\Psi^+)^{\otimes2},(\Psi^-)^{\otimes2}\}$, all four states can be discriminated perfectly with LO ($P_\text{LO}=1$). For this, one measures $\sigma_Z\otimes \sigma_Z$ ($\sigma_X\otimes \sigma_X$) on the first (second) copy.

In comparion, the five states \eqref{bsm5states} obey $P_\text{G}=\frac{4}{5}$ since they are four-dimensional. We are interested in comparing the ratio $R=\frac{P_\text{t}}{P_\text{G}}$ with the corresponding value of $R=\frac{1}{2}$ for the four Bell states. Numerical optimisation over LO strategies with arbitrary post-processing of the outcomes gives $P_\text{LO}\approx 0.3946$, i.e.~the lower ratio $R\approx 0.493<1/2$. No increase was found with one-way LOCC but PPT measurements reach $P_\text{PPT}=2/5$. If two copies of the states \eqref{bsm5states} are available, the difference with the Bell basis is more significant. A semidefinite program gives a non-unit global discrimination probabiltiy $P_\text{G}\approx 0.9964$, which contrasts the unit success probability  of discriminating the two-copy Bell basis already with LO. Under PPT measurements \cite{Cosentino2013}, we compute  that $P_\text{PPT}=\frac{4}{5}$ while for LO we numerically find at best $P_\text{LO}\approx 0.7297$. Both these cases correspond to values of $R$ well below the case of the Bell basis.

\section{Randomised matching} We now consider the equiangular and maximally distinguishable states \eqref{bsm5states} in the context of a simple task in entanglement-assisted quantum communication, named  randomised matching in Ref.~\cite{Feng2023}. Consider that Alice and Bob are given independent inputs $x,y\in\{1,\ldots,5\}$ respectively. Bob is tasked with deciding whether the inputs match, i.e.~whether $x=y$, by outputting $b\in\{0,1\}$. The average  success probability of randomised matching is
\begin{equation}
p_\text{win}=\frac{1}{25}\sum_{x,y}p(b=\delta_{x,y}|x,y).
\end{equation}
To perform the task, Alice and Bob are permitted the same resources as in dense coding; a shared two-qubit state and one qubit of communication. We show that the geometry of the states in \eqref{bsm5states} is crucial for optimally performing the task. Firstly, we have obtained an upper bound by relaxing the scenario to the strictly stronger setting in which Alice can  send four-dimensional states without shared entanglement, and then bounding the figure of merit using the semidefinite programing methods of \cite{Navascues2015, Rosset2019}. This gives $p_\text{win}\leq 19/20$. We will now construct a protocol in which this is saturated. To this end, let the shared state be $\ket{\phi^+}$  and let Alice perform the five unitaries in Appendix~\ref{AppUnitaries} associated with the states \eqref{bsm5states}. Using that $M_{0|y}+M_{1|y}=\openone$, the success probability can be re-written as $p_\text{win}=4/5+1/25\sum_y \tr(M_{1|y}V_y)$ where $V_y=\phi_y-\sum_{x\neq y} \phi_x$, where $\phi_x=\ketbra{\phi_x}$ is the global state after Alice's encoding, $\ket{\phi_x}=U_x\otimes \openone\ket{\phi^+}$, namely the states in Eq.~\eqref{bsm5states}. We then have $V_y=2\phi_y-\sum_x \phi_x=2\phi_y-\frac{5}{4}\openone$. Optimally choose Bob's measurement projectors as $M_{1|y}=\phi_y$; his five measurements thus correspond to partial Bell analysers of the states \eqref{bsm5states}. This gives $p_\text{win}=4/5+1/25\sum_y (2-5/4)=19/20$. This optimal quantum strategy is made possible by the existence of a non-projective BSM. Notably, it even outperforms classical strategies based on doubling the channel capacity \cite{Piveteau2022}, i.e.~sending two bits, since then $p_\text{win}\leq 23/25$ \cite{Feng2023}.

\section{Discussion} We have seen that non-projective Bell state measurements can be defined in a natural way through equiangular tight frames of maximally entangled states, and that such measurements are indeed possible in quantum theory. Our work focused on the case of two-qubit measurements, where we  determined both the largest possible BSM and the limitations relevant for a larger number of equally spaced maximally entangled states. However, it is not straightforward how many outcomes non-projective BSMs can have for bipartite systems of local dimension larger than two. Even determining in general whether it is possible to have one extra outcome, i.e.~$N=d^2+1$, appears non-trivial. In particular, when $d>2$, we can no longer rely on an isomorphism with Euclidean spaces but can still view the task as a geometry problem over $\text{SU}(d)$. Another natural open problem is to determine bounds, analogous to the Gerzon bounds, on the largest number of outcomes possible in a $d$-dimensional non-projective BSM. This is expected to be more challenging, since the set of maximally entangled states is not a vector space. A related basic problem is to bound the largest number of maximally entangled states that admit the equiangular property ($\alpha<1$). 

\begin{acknowledgments}
	We thank Ingemar Bengtsson for interesting discussions and Martin Renner for feedback. This work is by the Wenner-Gren Foundation, by the Knut and Alice Wallenberg Foundation
	through the Wallenberg Center for Quantum Technology (WACQT) and the Swedish Research Council under Contract No. 2023-03498.
\end{acknowledgments}

\bibliography{references_equiangular}

\appendix

\section{Proof of Welch bound}\label{AppWelch}
The Welch bound \cite{Welch1974} states that for any set of states $\{\ket{\psi_i}\}_{i=1}^N$ with $\ket{\psi_i}\in \mathbb{C}^m$ and $N \geq m$, it holds that 
\begin{equation}
\max_{i\neq j} \left|\braket{\psi_i}{\psi_j}\right|^2 \geq \alpha^*,
\end{equation}
with equality if the associated frame operator satisfies $S=\frac{N}{m}\openone$. Here, $\alpha^*=\frac{N-m}{m(N-1)}$. To search for ETFs, we look for sets of equiangular lines which meet this lower bound.

Consider the Gram matrix $G_{i,j}=\braket{\psi_i}{\psi_j}$. By construction, we have $G\succeq 0$  and $\tr(G)=N$. Since the states are $m$-dimensional, it holds that $\rank(G)\leq m$. Assign the non-zero eigenvalues of $G$ to the vector $\vec{\lambda}=(\lambda_{1},...,\lambda_{m})$, and define the normalized constant vector $\hat{u}=\frac{1}{\sqrt{m}}\textbf{1}_{m}$. Apply the Cauchy-Schwarz inequality
\begin{equation}
\bigg(\sum_{i=1}^{m}u_{i}\lambda_{i}\bigg)^{2} \le \bigg(\sum_{i=1}^{m}u_{i}^{2}\bigg)\bigg(\sum_{i=1}^{m}\lambda_{i}^{2}\bigg)
\end{equation} to obtain
\begin{equation}\label{eq:cauchyschwarz}
\bigg(\sum_{i=1}^{m}\lambda_{i}\bigg)^{2} \le m\sum_{i=1}^{m}\lambda_{i}^{2},
\end{equation} 
where the left hand side is $\tr(G)^{2}=N^{2}$. From the Frobenius norm of the Gram matrix $G$ denoted $\|G\|_{F}$, we have that 
\begin{equation}\label{eq:frobenius}
\|G\|_{F}^{2}=\sum_{i=1}^{N}\lambda_{i}^{2}=\sum_{i,j=1}^{N}|\langle  \psi_{i}|\psi_{j}\rangle|^{2}
\end{equation}
Substituting \eqref{eq:frobenius} into \eqref{eq:cauchyschwarz} and subtracting from both sides the $N$ terms where $i=j$,
\begin{equation}\label{eq:crossnormsquaredsum}
\sum_{i\ne j}^{N}|\langle \psi_{i}|\psi_{j}\rangle|^{2}\ge \frac{N(N-m)}{m}
\end{equation}
Since the average of a set of non-negative numbers cannot be greater than the largest in the set $c^{2}_{\max}$, divide both sides by $N(N-1)$, the number of terms in the sum, to obtain
\begin{equation}\label{stepp}
c^{2}_{\max} \ge \frac{1}{N(N-1)}\sum_{i\ne j}|\langle \psi_{i}|\psi_{j}\rangle |^{2} \ge \frac{N-m}{m(N-1)}.
\end{equation}
Finally, we obtain the lower bound for the maximum overlap squared for any given set of states
\begin{equation}
\alpha^*=c^{2}_{\max} \ge \frac{N-m}{m(N-1)}
\end{equation}
The lower bound is reached when equality is achieved in  both \eqref{eq:cauchyschwarz} and \eqref{stepp}. The former equality occurs when $\vec{\lambda}$ is proportional to $\hat{u}$. In other words, all non-zero eigenvalues of $G$ are equal to some constant, say $t$. Then, \eqref{eq:cauchyschwarz} gives $N^{2}=m^{2}t^{2}$ and $t=\frac{N}{m}$. Recall that $S$ and $G$ have the same eigenvalues so $S=\frac{N}{m}\openone$ when the Welch bound is saturated. The latter inequality occurs when  $|\langle \psi_{i}|\psi_{j}\rangle |$ is constant for all $i\neq j$. Taken together, this implies an ETF.

\section{Six equiangular maximally entangled states from the icosahedron}\label{AppIcosa}
In \cite{Lemmens1973}, the six non-antipodal vertices of the regular icosahedron are used to construct a set of six equiangular states on  $\mathbb{R}^3$. This can be trivially lifted also  to $\mathbb{R}^4$. These vectors are
\begin{align}
\ket{\psi_{1,2}} \sim
\begin{pmatrix}
0 \\ 1 \\ \pm q \\ 0
\end{pmatrix},  \quad 
\ket{\psi_{3,4}} \sim
\begin{pmatrix}
-1 \\ \pm q \\ 0 \\ 0
\end{pmatrix},  \quad
\ket{\psi_{5,6}} \sim
\begin{pmatrix}
\pm q \\ 0 \\ 1 \\ 0
\end{pmatrix},
\end{align}
where $q=\frac{1+\sqrt{5}}{2}$ is the golden ratio. Above, we have neglected the normalisation. After normalisation, the overlap constant $\alpha$ is $\alpha=\frac{q\sqrt{5}}{4}\approx 0.905$. 

Using the mapping from $\mathbb{R}^4$ to maximally entangled two-qubit states, given in the main text, we obtain the following set of equiangular maximally entangled states
\begin{align}\nonumber \label{eq:icosahedron}
&\ket{\psi_{1,2}}=\frac{1}{\sqrt{2}}\left(\Psi^+\mp iq \Psi^-\right), \\ \nonumber
&\ket{\psi_{3,4}}=\frac{1}{\sqrt{2}}\left(-i\Phi^+\pm q \Psi^+\right), \\ 
&\ket{\psi_{5,6}}=\frac{1}{\sqrt{2}}\left(\pm q\Psi^+- \Psi^-\right),
\end{align}
where $ \Phi^\pm=\frac{1}{\sqrt{2}}\left(\ket{00}\pm \ket{11}\right)$ and  $\Psi^\pm=\frac{1}{\sqrt{2}}\left(\ket{01}\pm \ket{10}\right)$. However, the overlap constant is far from saturating the Welch bound, which corresponds to $\alpha=\alpha^*=\frac{1}{10}$.

\section{Unitary representation of five-outcome  BSM}\label{AppUnitaries}
The five single-qubit unitaries corresponding to the five-outcome BSM are 
\begin{align}\nonumber
&U_1=\frac{i}{\sqrt{2}}\left(
\begin{array}{cc}
1 & -1 \\
1 & 1 \\
\end{array}
\right),\\\nonumber
& U_2=\frac{1}{\sqrt{2}}\left(
\begin{array}{cc}
d+ia & b-ic \\
b+ic & -d+ia \\
\end{array}
\right),\\\nonumber
& U_3=\frac{1}{\sqrt{2}} \left(
\begin{array}{cc}
-d+ia & -b-i c \\
-b+ic & d+ia \\
\end{array}
\right),\\\nonumber
& U_4=\frac{1}{\sqrt{2}}\left(
\begin{array}{cc}
b+i c & -d-ia \\
-d+i a & -b+ic \\
\end{array}
\right),\\
& U_5=\frac{1}{\sqrt{2}}\left(
\begin{array}{cc}
-b+ic & d-i a \\
d+ia & b+i c \\
\end{array}
\right),
\end{align}
%\begin{align}
%	\begin{split}
%		|\phi_{1}\rangle &=  |\Phi^{+}\rangle \\
%		|\phi_{2}\rangle &= -\frac{1}{4}|\Phi^{+}\rangle+\frac{\sqrt{5}}{4}|\Psi^{-}\rangle+\frac{i(b-d)}{2}|\Phi^{-}\rangle-\frac{i(b+d)}{2}|\Psi^{+}\rangle \\
%		|\phi_{3}\rangle &= -\frac{1}{4}|\Phi^{+}\rangle+\frac{\sqrt{5}}{4}|\Psi^{-}\rangle-\frac{i(b-d)}{2}|\Phi^{-}\rangle+\frac{i(b+d)}{2}|\Psi^{+}\rangle \\
%		|\phi_{4}\rangle &= -\frac{1}{4}|\Phi^{+}\rangle-\frac{\sqrt{5}}{4}|\Psi^{-}\rangle-\frac{i(b+d)}{2}|\Phi^{-}\rangle-\frac{i(b-d)}{2}|\Psi^{+}\rangle \\
%		|\phi_{5}\rangle &= -\frac{1}{4}|\Phi^{+}\rangle-\frac{\sqrt{5}}{4}|\Psi^{-}\rangle+\frac{i(b+d)}{2}|\Phi^{-}\rangle+\frac{i(b-d)}{2}|\Psi^{+}\rangle
%	\end{split}
%\end{align}

The constants are 
\begin{align}\nonumber
& a = \frac{-1+\sqrt{5}}{4} && b = \sqrt{\frac{5+\sqrt{5}}{8}}\\
& c = -\frac{1+\sqrt{5}}{4} &&  d = \sqrt{\frac{5-\sqrt{5}}{8}}.
\end{align}

\section{Five-outcome equiangular product measurement}\label{AppProduct}
The local unitaries corresponding to an equiangular five-outcome product measurement are 
\begin{align}\nonumber
\openone &\otimes \openone \\\nonumber
\left(\cos\frac{2\pi}{5}\openone+i\sin\frac{2\pi}{5}\sigma_X\right) &\otimes \left(\cos\frac{4\pi}{5}\openone+i\sin\frac{4\pi}{5}\sigma_X\right)\\\nonumber
\left(\cos\frac{3\pi}{5}\openone+i\sin\frac{3\pi}{5}\sigma_X\right)&\otimes \left(\cos\frac{\pi}{5}\openone+i\sin\frac{\pi}{5}\sigma_X\right)\\\nonumber
\left(\cos\frac{\pi}{5}\openone-i\sin\frac{\pi}{5}\sigma_X\right)&\otimes\left(\cos\frac{2\pi}{5}\openone-i\sin\frac{2\pi}{5}\sigma_X\right)\\
\left(\cos\frac{\pi}{5}\openone+i\sin\frac{\pi}{5}\sigma_X\right)&\otimes \left(\cos\frac{2\pi}{5}\openone+i\sin\frac{2\pi}{5}\sigma_X\right).
\end{align}

\section{Optimal six equiangular maximally entangled states}\label{AppUnitaries2}
The rotation angles can be chosen as 
\begin{align}\nonumber
& \gamma_0=0, & \gamma_1=\pi-\arccos\left(1/3\right),\\ \nonumber
& \gamma_2=\arccos\left(1/3\right), & \gamma_3=-\arccos\left(1/3\right),\\
& \gamma_4=\arccos\left(1/3\right), & \gamma_5=-\arccos\left(1/3\right).
\end{align}
The Bloch vectors can be chosen as the six respective columns of the following matrix 
\begin{equation}
\begin{pmatrix}\vspace{2mm}
0 & 0 & -\frac{\sqrt{15}}{4} & -\frac{1}{2}\sqrt{\frac{3}{5}} &  -\frac{1}{2}\sqrt{\frac{3}{5}} &  -\frac{3}{4}\sqrt{\frac{3}{5}}\\ \vspace{2mm}
0 & 0 & 0 & - \sqrt{\frac{3}{5}} &  -\sqrt{\frac{3}{5}} &  \sqrt{\frac{3}{5}}\\
0 & 1 & -\frac{1}{4} & -\frac{1}{2} & \frac{1}{2} & \frac{1}{4}
\end{pmatrix}.
\end{equation}
It is straightforward to verify that the equiangular condition holds with $\alpha=\frac{1}{9}$.

\section{Characteristic polynomials of Gram matrix}\label{AppPolynomials}
We consider the set of $6\times 6$ Gram matrices with $\pm 1/\sqrt{\alpha}$ on the off-diagonal entries to be equivalent up to spectrum-preserving row and column operations described in the text. It was determined in \cite{vanLint1966} that there are 16 distinct classes of Gram matrices, and accordingly, 16 distinct characteristic polynomials. They are computed in Mathematica and are listed in order corresponding to the graphs in \cite{vanLint1966} from left to right:
\begin{widetext}
	\begin{align}
	\begin{split}
	p_{1}(\mu,\lambda)&= (\lambda -5\mu-1)(\lambda +\mu -1)^{5} \\
	p_{2}(\mu,\lambda)&=(\lambda-\mu-1)(\lambda+\mu-1)^{3}(\lambda^{2}-2(1+\mu)\lambda - 11\mu^{2}+2\mu+1) \\
	p_{3}(\mu,\lambda) &= (\lambda+\mu-1)^{3}(\lambda^{3}-3(1+\mu)\lambda^{2}+3(1+2\mu-3\mu^{2})\lambda +19\mu^{3}+9\mu^{2}-3\mu-1) \\
	p_{4}(\mu,\lambda)&=(\lambda+\mu-1)(\lambda^{2}-2\lambda-5\mu^{2}+1)(\lambda^{3}-(3+\mu)\lambda^{2}+(3+2\mu-9\mu^{2})\lambda+\mu^{3}+9\mu^{2}-\mu-1) \\
	p_{5}(\mu,\lambda)&= (\lambda-\mu-1)^{2}(\lambda+\mu-1)(\lambda+3\mu-1)(\lambda^{2}-2(1+\mu)\lambda-7\mu^{2}+2\mu+1) \\
	p_{6}(\mu,\lambda)&=(\lambda-3\mu-1)^{2}(\lambda+\mu-1)^{3}(\lambda+3\mu-1)\\
	p_{7}(\mu,\lambda) &= (\lambda-\mu-1)(\lambda^{2}-2\lambda-5\mu^{2}+1)(\lambda^{3}-(3-\mu)\lambda^{2}-(9\mu^{2}+2\mu-3)\lambda-\mu^{3}+9\mu^{2}+\mu-1)\\
	p_{8}(\mu,\lambda) &= ((\lambda-1)^{2}-\mu^{2})^{2}((\lambda-1)^{2}-13\mu^{2}) \\
	p_{9}(\mu,\lambda) &= (\lambda-\mu-1)(\lambda-3\mu-1)(\lambda+\mu-1)(\lambda+3\mu-1)(\lambda^{2}-2\lambda -5\mu^{2}+1) \\
	p_{10}(\mu,\lambda) &= (\lambda-\mu-1)(\lambda+\mu-1)(\lambda^{4}-4\lambda^{3}+(6-14\mu^{2})\lambda^{2}+(28\mu^{2}-4)\lambda+29\mu^{4}-14\mu^{2}+1) \\
	p_{11}(\mu,\lambda) &=(\lambda-\mu-1)^{3}(\lambda-3\mu-1)(\lambda+3\mu-1)^{2} \\
	p_{12}(\mu,\lambda)&= (\lambda-\mu-1)(\lambda-3\mu-1)(\lambda+\mu-1)^{2}(\lambda^{2}+2(\mu-1)\lambda-7\mu^{2}-2\mu+1) \\
	p_{13}(\mu,\lambda) &= (\lambda-\mu-1)(\lambda^{2}-2\lambda-5\mu^{2}+1)(\lambda^{3}+(\mu-3)\lambda^{2}-(2\mu+9-3)\lambda-\mu^{3}+9\mu^{2}+\mu-1) \\
	p_{14}(\mu,\lambda) &= (\lambda-\mu-1)^{3}(\lambda^{3}+3(\mu-1)\lambda^{2}-3(3\mu^{2}+2\mu-1)\lambda-19\mu^{3}+9\mu^{2}+3\mu-1) \\
	p_{15}(\mu,\lambda)&=(\lambda-\mu-1)^{3}(\lambda+\mu-1)(\lambda^{2}+2(\mu-1)\lambda-11\mu^{2}-2\mu+1) \\
	p_{16}(\mu,\lambda) &= (\lambda-\mu-1)^{5}(\lambda+5v-1)
	\end{split}
	\end{align}
\end{widetext}

The Gram matrix has rank $r \le 4$ if it has eigenvalue $\lambda=0$ with multiplicity $6-r$, so $\mu$ must be chosen so that $\lambda^{6-r}$ is a factor of the characteristic polynomial.  Impose $|\mu| < 1$, so that the set of equiangular lines is not trivial. For example, the characteristic polynomial $p_{1}$ does not correspond to the Gram matrix of a set of equiangular lines since $\mu=1/5$ and $\mu=1$ produce zero eigenvalues, but one corresponds to a Gram matrix with rank $r=5$ and the other produces trivial equiangular lines. For this reason also, $p_{2}$ is infeasible since only $\mu=\pm 1$ produce $\lambda=0$ as eigenvalues. Proceeding in this way for each of the 16 characteristic polynomials, we find that $p_{13}$ admits a solution for $\mu=1/\sqrt{5}$, and $p_{6}$ and $p_{11}$ permit solutions for $\mu=-1/3$ and $\mu=1/3$ respectively. $p_{13}$ is precisely the characteristic polynomial of the Gram matrix of the six equiangular lines in \eqref{eq:icosahedron}. We see that $\mu=\pm 1/3$, which implies $\alpha=1/9$, is the smallest value which satisfies the rank and positivity requirements of the Gram matrix so that a decomposition $G=V^{\dagger}V$ can be made corresponding to six equiangular lines.

\end{document}